\documentclass[fleqn,10pt]{wlscirep}
\usepackage{graphicx}
\usepackage{subfigure}
\usepackage[utf8]{inputenc}
\usepackage[T1]{fontenc}
\usepackage{float}
\usepackage{color}
\usepackage{amsmath,amsthm,amssymb}
\usepackage{hyperref}
\usepackage{lineno}

\title{City-scale synthetic individual-level vehicle trip data}

\author[1]{Guilong Li}
\author[1]{Yixian Chen}
\author[1]{Yimin Wang}
\author[4]{Peilin Nie}
\author[3]{Zhi Yu}
\author[1,2*]{Zhaocheng He}
\affil[1]{Guangdong Provincial Key Laboratory of Intelligent Transportation System, School of Intelligent Systems Engineering,
Sun Yat-sen University, Guangzhou, 510275, China}
\affil[2]{The Pengcheng Laboratory, Shenzhen 518055, China}
\affil[3]{Joint Research and Development Laboratory of Smart Policing in Xuancheng Public Security, Xuancheng, 242000, China}
\affil[4]{School of Environmental and Chemical Engineering,
FoShan University, FoShan, 528000, China}
\affil[*]{corresponding author: Zhaocheng He (hezhch@mail.sysu.edu.cn)}

\begin{abstract}

Trip data that records each vehicle's trip activity on the road network describes the operation of urban traffic from the individual perspective, and it is extremely valuable for transportation research. However, restricted by data privacy, the trip data of individual-level cannot be opened for all researchers, while the need for it is very urgent. In this paper, we produce a city-scale synthetic individual-level vehicle trip dataset by generating for each individual based on the historical trip data, where the availability and trip data privacy protection are balanced. Privacy protection inevitably affects the availability of data. Therefore, we have conducted numerous experiments to demonstrate the performance and reliability of the synthetic data in different dimensions and at different granularities to help users properly judge the tasks it can perform. The result shows that the synthetic data is consistent with the real data (i.e., historical data) on the aggregated level and reasonable from the individual perspective.

\end{abstract}

\begin{document}

\flushbottom
\maketitle
\thispagestyle{empty}

\noindent Key words: Urban transportation, synthetic individual-level trip data, data privacy protection, trip generation.

\section*{Background \& Summary}

With the popularity of data-driven methods, data has become the foundation for urban transportation research today. Although there are some datasets\cite{du2018temporal,du2019inter,lai2022global} that represent human mobility have been opened, they have limited benefit for solving transportation issues since these data are collected by non-transportation activities and cannot be interpreted as trip behaviour directly\cite{zhao2018individual}. Hence, the data directly obtained from the transportation system is necessary for transportation researches. 

In the past, limited by the capability of detectors, only the traffic data at the aggregated level, like volume, could be obtained. These data characterize traffic conditions in different dimensions, based on which plenty of 
studies\cite{okutani1984dynamic,hamed1995short,zhu2014traffic} have been developed to assist in traffic management. However, traffic condition is formed by trip activities of individuals, which is not contained, and only statistical values about them remain in aggregated data. Thus, such data cannot support travellers' behaviour mining\cite{KUSAKABE2014179}, personalized transportation guidance\cite{kuhail2018smart}, and other individual trip studies\cite{li2018multi,chen2019simulation,cheng2021incorporating,li2022potential}, which are in high demand for refined traffic management nowadays. In this case, the data indicating individual trip activities is urgently needed, which we call individual-level trip data. Individual-level trip data describes the micro-operation of urban traffic system, and it contains each individual's trip information, including trip time, origin, destination, and path. Aggregated-level traffic data can be obtained from it by counting, so individual-level trip data can also support studies using aggregated data. 

Individual-level trip data is now available through identity detection devices with data processing like trajectory reconstruction\cite{wang2023city}. Although it is technically accessible, the individual-level trip data is extremely difficult to obtain and open for two reasons. First, individual-level trip data collection is expensive and restricted by local policies, which leads to only a few researchers who have cooperated with the government can get it. The second is that the real individual-level trip data involves data privacy that has been discussed by researchers\cite {gao2019quantifying}, making it almost impossible to share. For this reason, studies that used individual-level trip data cannot open their dataset\cite{zhao2018individual,rao2018origin,sun2021joint,gao2019quantifying}. So, there are high demands for individual-level trip data, but it is hard to access.

In this paper, we will propose a city-scale synthetic individual-level trip dataset containing 1,829,218 trip records of 276,978 vehicle individuals in Xuancheng for one week. Each record of the dataset represents one trip of an individual, containing departure time at the minute level, trip origin, and destination represented by traffic zone, as well as the trip path that consists of a sequence of roads. Besides, because of our data-mining works, there is a field to indicate the traveller type of the individual, like commuter.

Unlike removing sensitive records that will cause statistical bias, we developed an individual trip generation method that balances data availability with the protection of individual trip privacy. With the historical trips input, it can generate trips for each individual and finally form this synthetic dataset. In terms of aggregate metrics like trip time distribution and the frequency of different origin-destinations of all individuals, the synthetic data is highly consistent with the real data. However, the synthetic dataset is not precisely aligned with the real data for trip privacy protection. Still, the generated trips of each individual are reasonable and can support research and analysis on the individual level. 

The synthetic individual-level vehicle trip dataset has a wide range of use for research, such as studies focusing on travellers' trip behaviour pattern\cite{chen2016promises}, trip prediction\cite{sun2021joint,jiang2021dp,9781338}, travel time estimation\cite{li2018multi,ramezani2012estimation}, origin-destination pattern estimation\cite{rao2018origin} and analysis of the effect of transportation policies\cite{liu2018effects}. Also, this dataset can support studies using aggregated data like traffic volume\cite{tang2021traffic,shao2018license}. Besides, since the road network data that matches this trip dataset is opened by this paper, simulation-based transportation research\cite{javid2018framework,chen2019simulation,ahn2008effects,hou2019adaptive} can also be supported by this dataset.


\section*{Methods}

\subsection*{Original data sources}

\begin{figure}[H]
\centering
\subfigure{\includegraphics[width=0.80\textwidth]{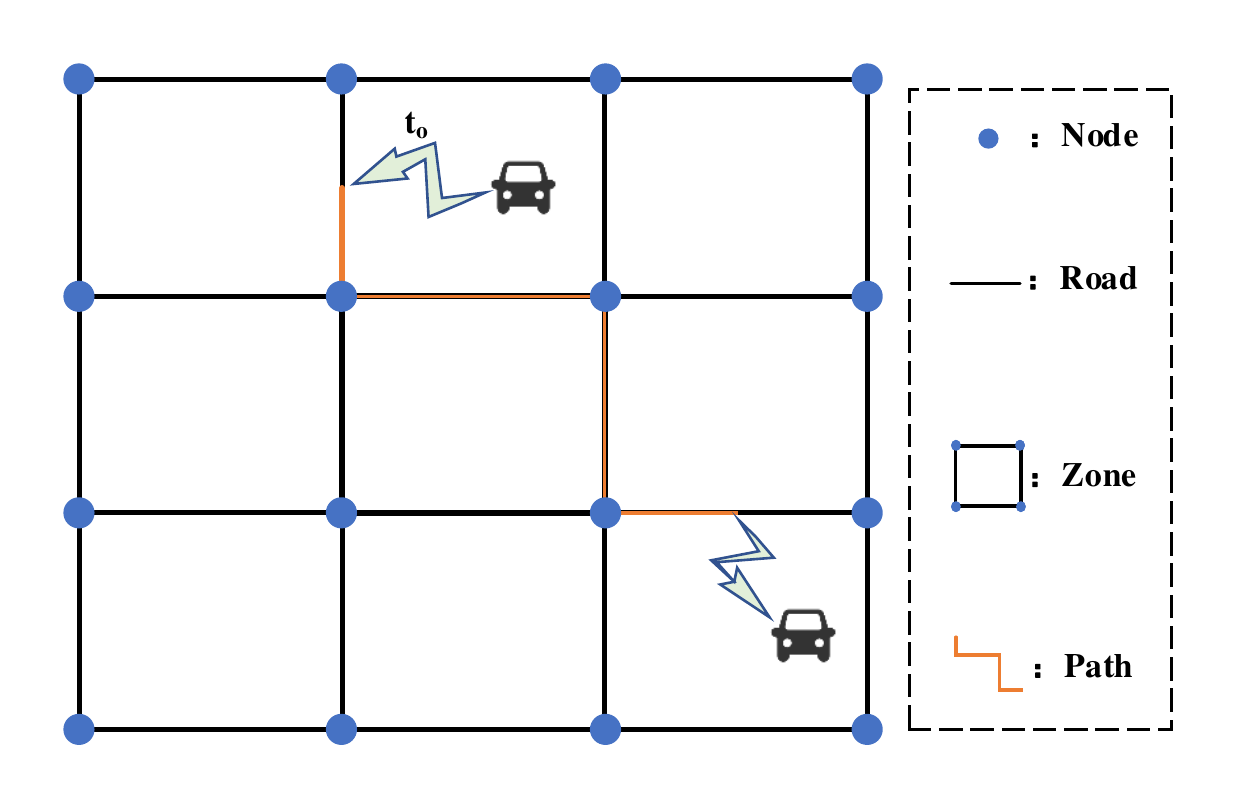}}
\caption{Schematic illustration of elements in road networks.}
\label{fig:road_zone}
\end{figure}

The mobility of vehicles on urban road networks can be captured by Automatic Vehicle Identification (AVI) \cite{bernstein1993automatic} device. With technologies like trajectory reconstruction\cite{wang2023city}, the data directly recorded by AVI device can be processed as individual-level trip data, which is more user-friendly and valuable. Specifically, each record of it covers information about one trip taken by an individual, including departure time, origin, destination, and trip path. Fig. \ref{fig:road_zone} gives a simple example of a road network in a regular square grid, and some elements are shown on it.

\begin{figure}[H]
\centering
\subfigure{\includegraphics[width=0.90\textwidth]{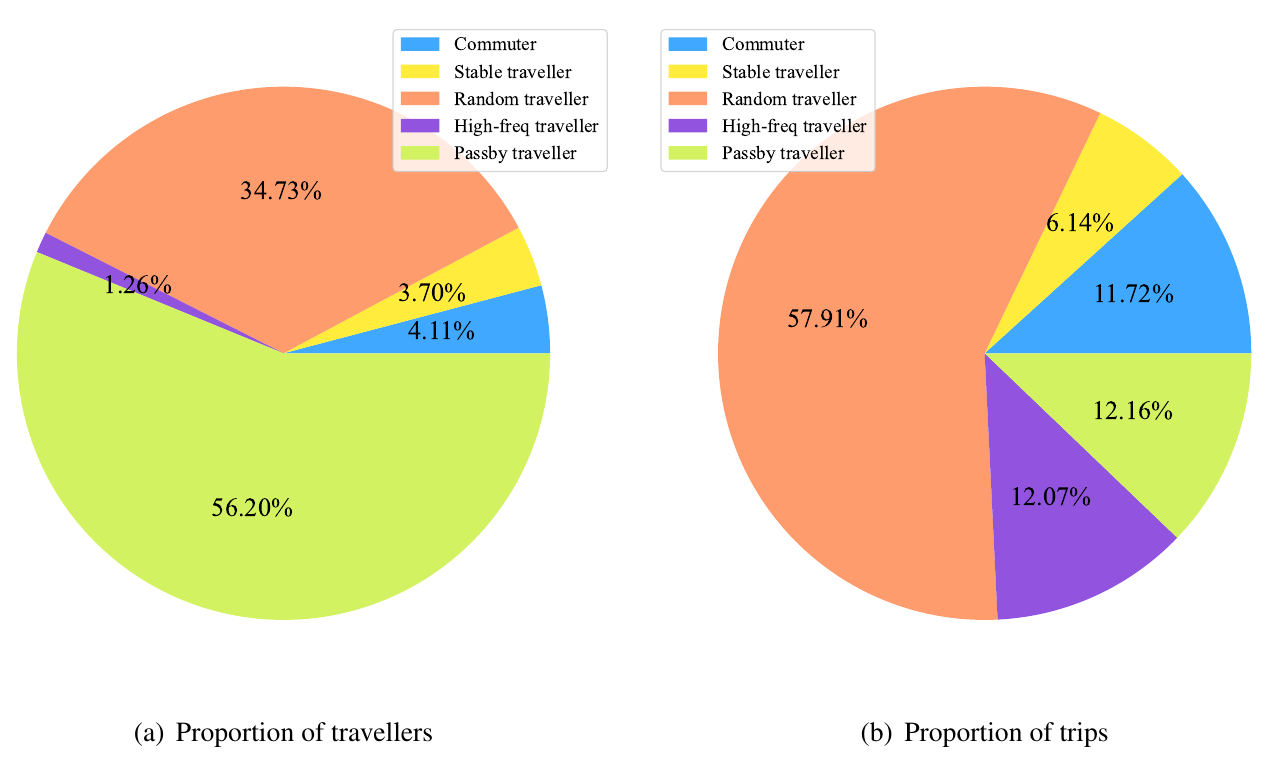}}
\caption{Proportion of travellers and trips of different traveller types.}
\label{fig:veh_trip_pro}
\end{figure}

\begin{figure}[H]
\centering
\subfigure{\includegraphics[width=0.98\textwidth]{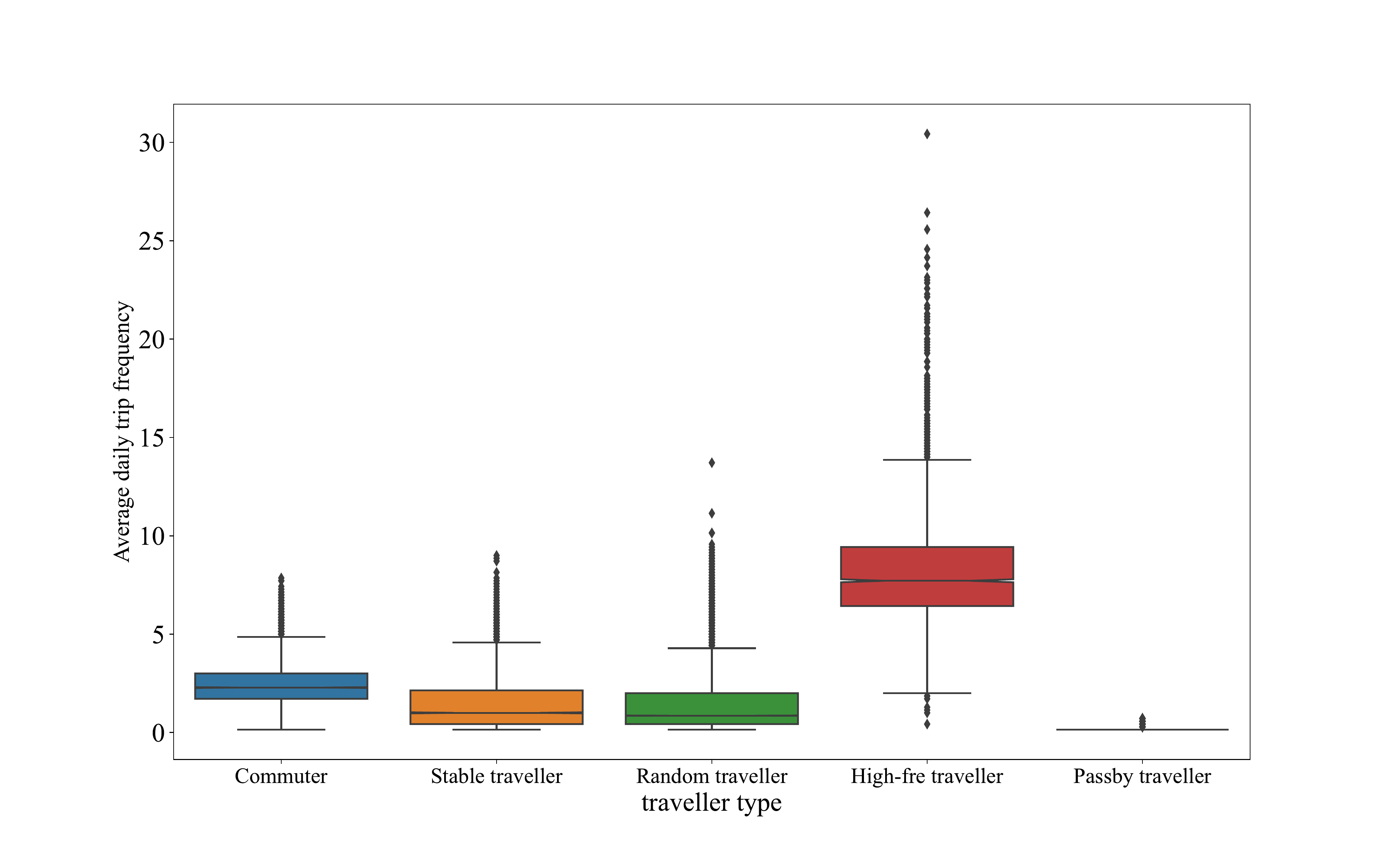}}
\caption{Average daily trip frequency distribution of different traveller types.}
\label{fig:veh_trip_fre}
\end{figure}

\begin{figure}[H]
\centering
\subfigure{\includegraphics[width=0.98\textwidth]{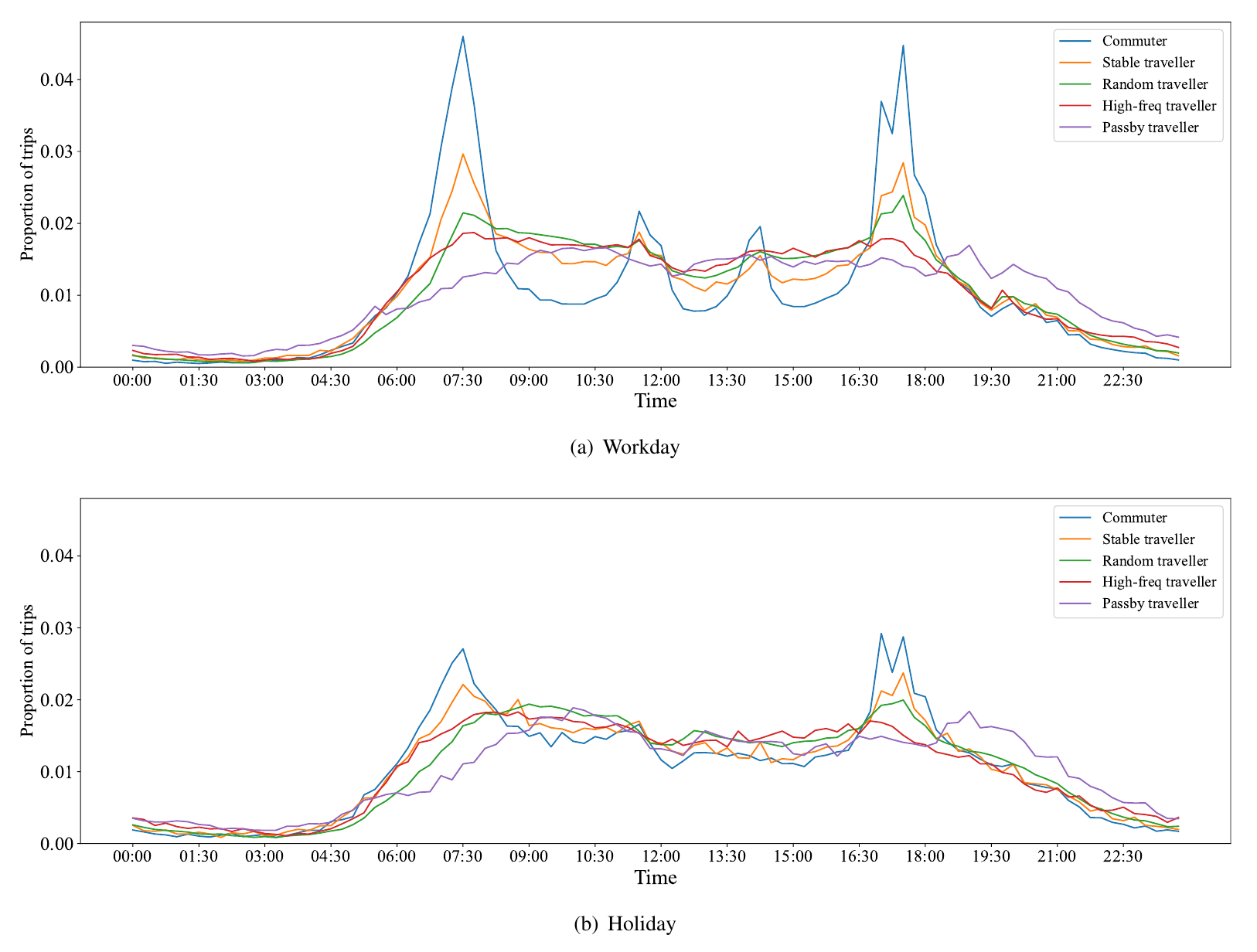}}
\caption{Distribution of trips with time of different traveller types (Statistics and plots at 15-minute granularity).}
\label{fig:veh_trip_time}
\end{figure}

\begin{figure}[H]
\centering
\subfigure{\includegraphics[width=0.98\textwidth]{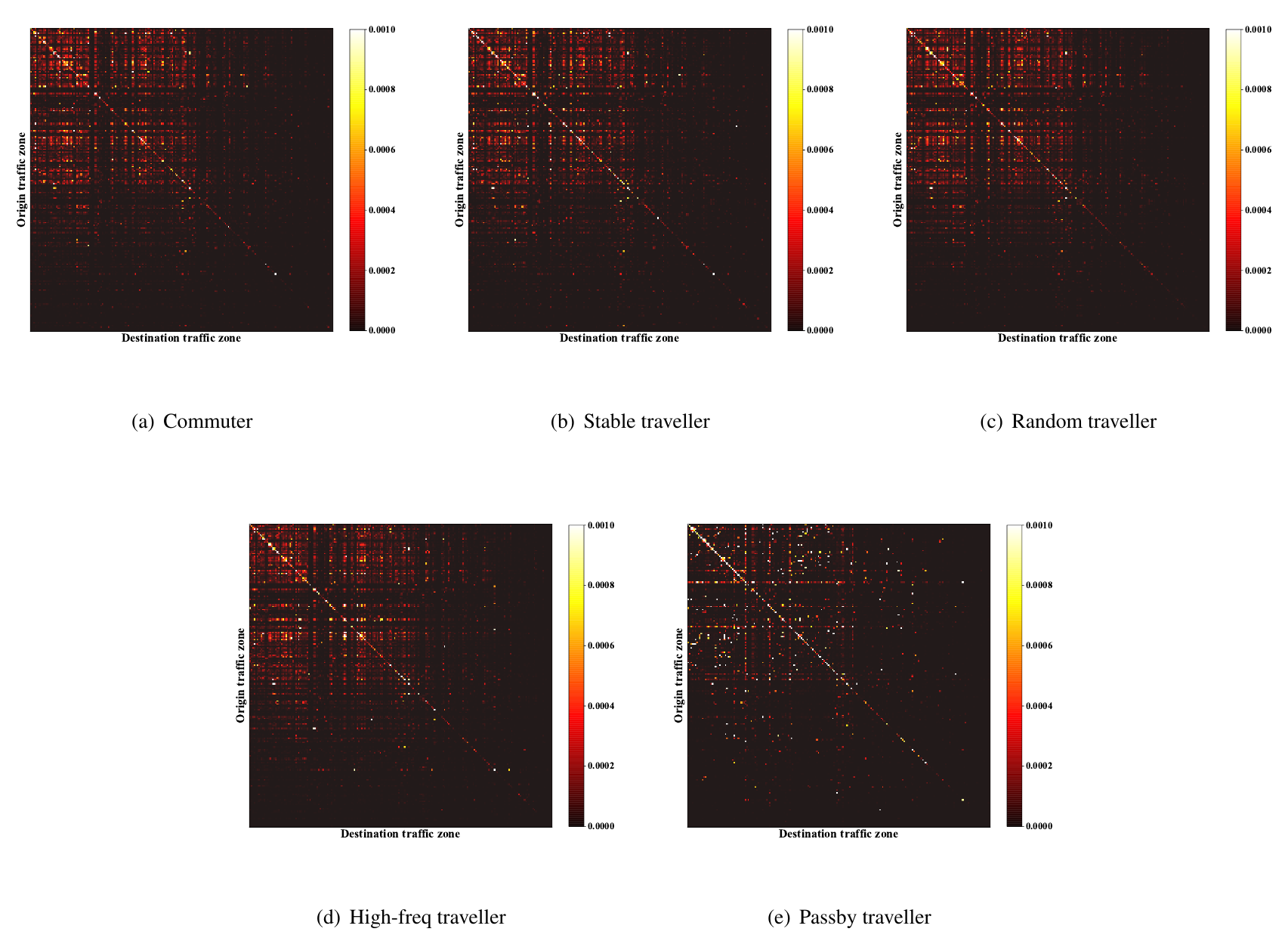}}
\caption{Trip distribution of different origin-destinations (ODs) of different traveller types.}
\label{fig:veh_trip_spa}
\end{figure}

\begin{figure}[H]
\centering
\subfigure{\includegraphics[width=0.92\textwidth]{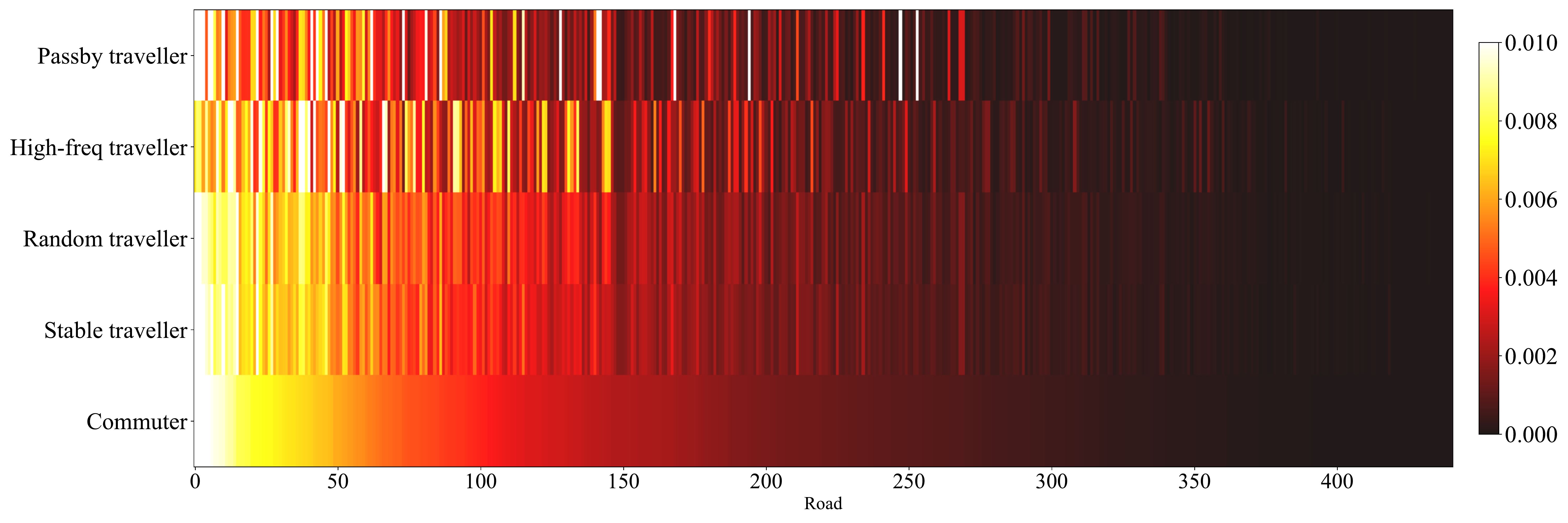}}
\caption{Road access frequency of different traveller types.}
\label{fig:veh_trip_flow}
\end{figure}

In this paper, the original trip data was collected in XuanCheng city of China for one month (2019/8/01-2019/9/02). It is a city-scale dataset containing 823,177 vehicles and 9,002,572 trips in total. In addition to the trip information mentioned above, this original trip dataset has two characteristics. First is that the trip origin and destination are represented by the traffic zone that is enclosed by roads (see Fig. \ref{fig:road_zone}).
In this case, the trip data is more reasonable and gets higher availability. Besides, some data mining works have been done on the dataset, by which each vehicle individual was given a traveller type like commuter. The proportion of travellers and trips of different traveller types are shown in Fig. \ref{fig:veh_trip_pro}. In addition, Fig. \ref{fig:veh_trip_fre}-\ref{fig:veh_trip_flow} show the trip information of travellers with different traveller types, of which the road access frequency refers to the proportion of trips through the roads (both directions will be counted).

\subsection*{Trip generation}

\begin{figure}[H]
\centering
\subfigure{\includegraphics[width=0.80\textwidth]{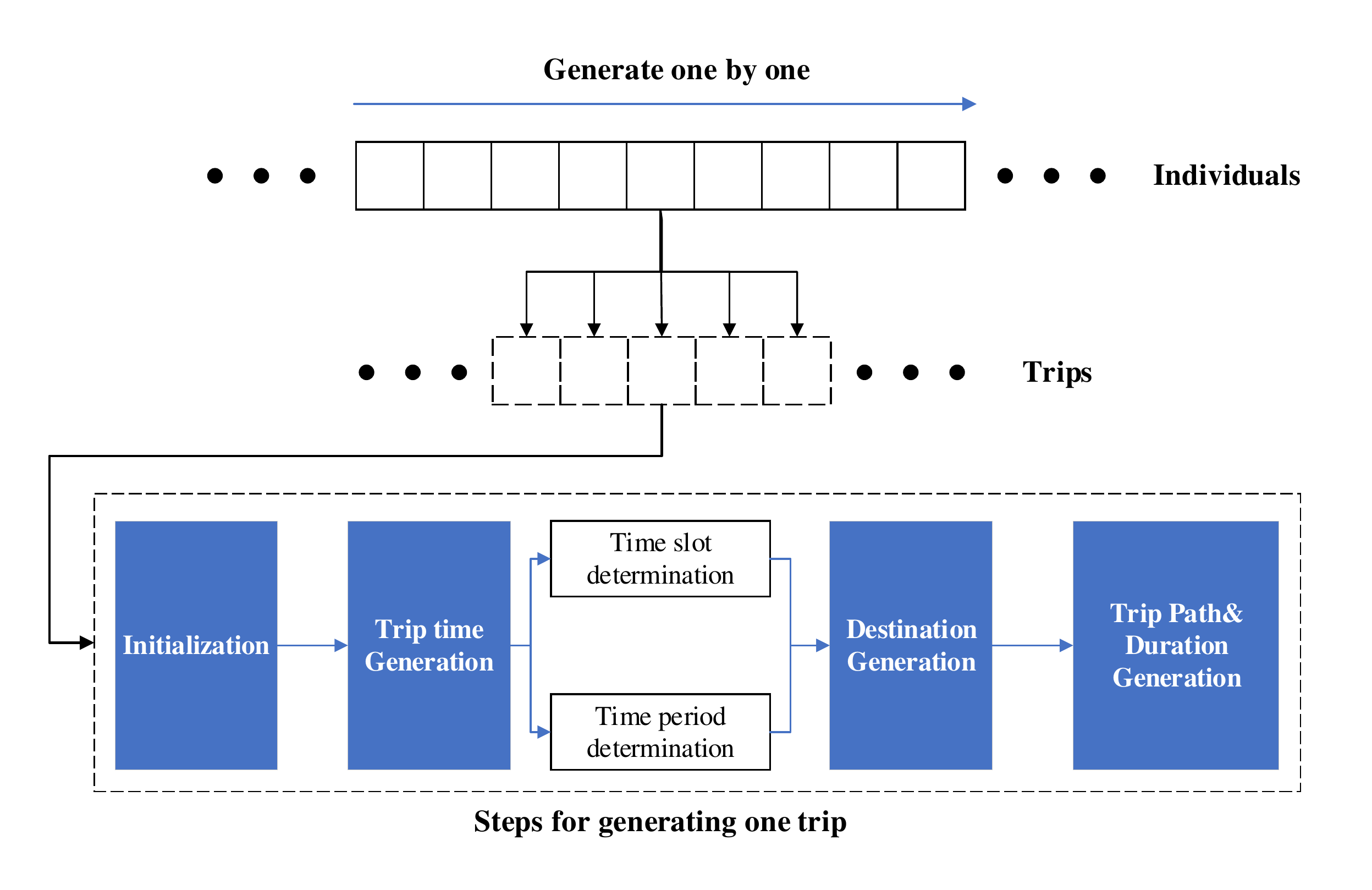}}
\caption{The framework of the trip generation.}
\label{fig:framework}
\end{figure}

As shown in Fig.\ref{fig:framework}, the individual's trips are generated one by one, and a new individual will be switched to when the former has completed generation. The generation for one trip can be decomposed into four steps. Before introducing each step, some definitions and notions need to be stated first.

Note the number of individuals in the original dataset as $n$, who are the targets for generation. Let $v_i$ be the $i$-th individual, then the set of individuals can be represented by $V=\{v_1, v_2, ..., v_n\}$. Some numeric variables about individuals are described in Table.\ref{tab:notations}. All of them are counted based on the range of the original dataset. Use $Z$ to represent the set of traffic zones in the city, and note $z_x \in Z$ as the traffic zone numbered $x$. Let $d_u$ be a specific day, and $d_{u+1}$ represents the day after $d_u$. Note the time of a day as $\{t_1,t_2, ..., t_{1440}\}$, of which $t_k$ represents a time period with an interval of one minute. For instance, $t_1$ indicates the time period ``$00:00-00:01$". Further, denote $l_{m,n}=\{t_m,t_{m+1}, ..., t_n \}$ as a time slot and the set of $l_{m,n}$ (i.e., $\{l_{1,a}, ..., l_{b,1440}\}$) as $L$. The time periods aggregated by a time slot are continuous, and each $t_k$ only belongs to one $l_{m,n}$ ($\exists!l_{m,n}:t_k \in l_{m,n},  \forall t_k$). 
$T^p$ is a variable maintained to indicate the current time of the generation, and it will be updated as the generation proceeds. The format of $T^p$ is $d_u\&t_k$, representing the time period ($t_k$) of the day ($d_u$). Note the end time of generation as $T^e$. It signs the generation of the individual is complete when $T^p>T^e$ and it's time to switch to a new one.

\subsubsection*{Initialization} \label{sec:ini}

Initialization is only executed when the switch is made during generation. Specifically, it can be classified into two cases. First, when switching to a new individual, an initial location (i.e., traffic zone) should be given as the origin of the first generated trip (after that, the destination of the last trip would be set as the origin of the next trip). Besides, the current time should be initialled as the start time of the generation. Next, when a new day is switched to, of course, it includes the first day of a new individual, the trip frequency of the individual for the day needs to be determined by initialization.

Note the initial location of $v_i$ as $z^p_i$. We set $z^p_i$ as the traffic zone that is most frequently visited by $v_i$, see Eq. \ref{eq:ini_loc}. 

\begin{equation} \label{eq:ini_loc}
    z^p_i=z_{\mathop{argmax}_{k}(v^o_{i,k}+v^d_{i,k})}
\end{equation}

Let $\overline {v}^f_i$ be the average of trip frequency per day of $v_i$, and it can be calculated by Eq. \ref{eq:trip_frc}, where $D$ is the number of days for $v^f_i$ counting. $\overline {v}^f_i$ can be represented as $\overline {v}^f_i=\left \lfloor \overline {v}^f_i \right \rfloor+\{\overline {v}^f_i\}$. On this basis, the number of trips needed to be generated of the day can be calculated by Eq. \ref{eq:fra_day}, where $\xi \sim B(1,\{\overline {v}^f_i\})$, i.e., $P\{\xi = k\} = {\{\overline {v}^f_i\}}^k(1-\{\overline {v}^f_i\})^{1-k}$, $k=0,1$. Note the number of trips that have been generated for $v_i$ of the day as $v^{dh}_i$. Then $\overline {v}^d_i - v^{dh}_i$ indicates the number of trips that remain to be generated. When $\overline {v}^d_i - v^{dh}_i=0$, a new day would be switched to, as well as recalculated $\overline {v}^d_i$. It is worth noting that, in this way, the single-day trip frequency of generated individuals will be more evenly distributed.

\begin{equation} \label{eq:trip_frc}
    \overline {v}^f_i = \frac{v^f_i}{D}  
\end{equation}

\begin{equation} \label{eq:fra_day}
    \overline {v}^d_i = \left \lfloor \overline {v}^f_i \right \rfloor + \xi
\end{equation}

\subsubsection*{Trip time generation}

The trip time mentioned in the section refers to the departure time of trips. Its granularity is at the minute level, which is equal to the original data. There are two steps to determine the trip time: 1) time slot determination; 2) time period determination.

\paragraph{Time slot determination} \label{sec:tsd}

This step determines the departure time slot of the trip being generated. An individual's trips should be spatially continuous, i.e., the destination of the previous trip should be the origin of the next trip after the individual's trips are ordered chronologically. To satisfy this objective fact of trips as much as possible, we generate trips of each traveller by time with $T^p$ recording the current time. In other words, the trip time of the last generated trip would be later than the former one. In this case, the present location of individuals is explicit, which benefits to guarantee spatial continuity. Besides, it is consistent with the law of individuals travelling in the real world. To achieve this, we introduce the logic factor $c^s$ of time slots.  

Suppose $T^p = d_r\&t_d$ and a trip in $d_r$ is being generated. Define the subsequent time slot set of $t_d$ as $L_s = \{l_{m,n} \in L|\exists m<k<n:\ k>d\}$. The time slot that $t_d$ belongs to is contained in $L_s$, enabling individuals to make multiple trips in the same time slot. Denote set $L_e \subseteq L_s$, and $L_e$ satisfies $|L_e| = min( \overline {v}^d_i-v^{dh}_i-1,|L_s|-1)$. If $|L_e| \neq 0$ (i.e., $|L_e| \neq \phi$), $|L_e|$ further follows the constraint: $\forall l_{m,n} \in L_e, l_{p,q} \in L_s-L_e: m>p$. Under the above constraints, $L_e$ contains the latest time slots of the day. Note set $A = L-L^c_s-L_e$ (i.e., $L_s-L_e$). 
On this basis, $\forall l_{m,n} \in L$, the logic factors can be calculated by Eq. \ref{eq:ts_der0}. $\kappa$ is a very small value, and its constraints will be described later. It can be proved that $A\neq \phi$. 


\begin{equation} \label{eq:ts_der0}
    c^s(l_{m,n}) = \left\{
    \begin{array}{l@{\qquad}l}
        1 & l_{m,n} \in A\\
        \kappa & else\\
    \end{array}
    \quad\qquad l_{m,n} \in L
    \right.
\end{equation}

The aggregated temporal distribution of trips portrays the urban traffic operation in the time dimension, which is valuable for transportation research\cite{thomas2008variations,hamed1995short,hou2019adaptive}. To restore the distribution of the real data and guarantee data availability at the aggregated level,  we first ensure the consistency of the proportion of trips among time periods. To achieve this, we introduce the aggregation factor $c^r$ of time slots.

Note the number of trips taken within $l_{m,n}$ by $v_i$ as $v^{l}_{i,m,n}$, i.e., $v^{l}_{i,m,n} = \sum^n_{k=m}(v^t_{i,k})$. Denote $u(x)$ as an aggregate function for individuals, and $u(l_{m,n}) = \sum_{i=1}^n(v^l_{i,m,n})$. Similarly, we denote $u^g(l_{m,n})$ for counting the generated data to measure the difference between generated and real data. Then the aggregation factor can be calculated by Eq. \ref{eq:ts_der1}, where $f(x)$ is a continuous function that satisfies the constraints shown in Eq. \ref{eq:fx}. 

\begin{equation} \label{eq:ts_der1}
    c^r(l_{m,n}) = f(u^g(l_{m,n})/\sum_{l_{m,n}\in L}u^g(l_{m,n}) - u(l_{m,n})/\sum_{l_{m,n}\in L}u(l_{m,n}))  \quad\qquad  l_{m,n} \in L
\end{equation}

\begin{equation} \label{eq:fx}
    f(x) \  satisfies.\left\{
    \begin{array}{l@{\qquad}l}
        f(x_1) > f(x_2) &,\ x_2>x_1\\
        f(0) = 1 & \\
        f(1) = 0 & \\
    \end{array}
    \quad\qquad
    x \in [-\mu,\mu]
    \right.
\end{equation}

As shown in Fig.~\ref{fig:veh_trip_time}, we have divided the travellers into five types based on our previous work. Their different distributions indicate they have different proportions of trips in time slots, such as the trips of commuters (during weekdays) are mainly concentrated in the morning and evening rush hours, which is important information for research. Therefore, to preserve the information for each type of traveller, we propose that $c^r$ only share among the same type of travellers when performing trip generation.

The larger the time granularity considered for the trip time, the weaker the uniqueness of individual temporal features. For instance, there may be only one individual trip at time period $t_q$, $t_w$ and $t_e$ in a day, but it will be hidden among many individuals considering the time slots these time periods belong to. 
Hence, for time slot choices of individuals, we make it consistent with one's choice in real data to get better usability on the individual level. Then two individual preference factors are designed: 1) whole preference factor; 2) location preference factor.

The whole preference factor $c^p$ of the time slot is defined by Eq. \ref{eq:ts_der2}.
Let $v^{l,o}_{i,m,n,a} = \sum^n_{k=m}(v^{t,o}_{i,k,a})$, which denotes the trip frequency of $v_i$ with $z_a$ as the origin during $l_{m,n}$. Suppose $z^p_i = z_c$, then location preference factor $c^{op}$ can be calculated by Eq. \ref{eq:ts_der2}.

\begin{equation} \label{eq:ts_der2}
    \begin{array}{l@{\qquad}l}
        c^p(l_{m,n}) = v^l_{i,m,n}/\sum_{l_{m,n}\in L}(v^l_{i,m,n}) & \\
        c^{op}(l_{m,n}) = v^{l,o}_{i,m,n,c}/\sum_{l_{m,n}\in L}(v^{l,o}_{i,m,n,c}) & \\
    \end{array}
    l_{m,n} \in L
\end{equation}

After defining $c^s(l_{m,n}),c^r(l_{m,n}),c^p(l_{m,n})$ and $c^{op}(l_{m,n})$, we give the formula for factor integration, see Eq. \ref{eq:ts_der3}. Then the probability of each time slot to be chosen is given by Eq. \ref{eq:ts_der4}.

\begin{equation} \label{eq:ts_der3}
c(l_{m,n}) = c^s(l_{m,n})*c^r(l_{m,n})*(c^p(l_{m,n})*(1+ c^{op}(l_{m,n}))+\varepsilon)
\end{equation}

\begin{equation} \label{eq:ts_der4}
p(l^s=l_{m,n}) = c(l_{m,n})/\sum_{l_(m,n) \in L} c(l_{m,n})
\end{equation}

Eq. \ref{eq:ts_der3} can be seen as three terms, and they consider trip time logic, aggregation information, and individual preference, respectively. $\varepsilon >0$ is a small value, and it ensures $c^p(l_{m,n})*(1+ c^{op}(l_{m,n}))+\varepsilon >0$, making a time slot will not be excluded only by individual preference.
On this basis, it can be proved that  $\sum_{l(m,n) \in L} c(l_{m,n}) > 0$. So whatever the case, Eq. \ref{eq:ts_der4} can pick a time slot with strong robustness. Conflicts may occur between different factors. For example, suppose a time slot $l_{i,o}$, its proportion in generation data is much lower than the real, i.e., $u^g(l_{i,o})/\sum_{l_{i,o}\in L}u^g(l_{i,o}) - u(l_{i,o})/\sum_{l_{i,o}\in L}u(l_{i,o}) \rightarrow -\mu$, but the individual never tripped on $l_{i,o}$ ($c^p(l_{i,o})=0$). This lets $c^r$ give a high value for balancing at aggregation, while $c^p(l_{m,n})*(1+ c^{op}(l_{m,n}))+\varepsilon = \varepsilon$ that a small value since $l_{i,o}$ is not preferred by the individual. To handle these conflicts, we determine the priority of these factors by following constraints.

\begin{equation} \label{eq:ts_der5}
    s.t. \left\{
    \begin{array}{l@{\qquad}l}
        1/\lim\limits_{x \to -\mu}f(x)\ll \varepsilon \ll 1/v^f_i & \\
        \kappa \lim\limits_{x \to -\mu}f(x) \approx 1 & \\
    \end{array}
    \right.
\end{equation}

$1/\lim\limits_{x \to -\mu}f(x)\ll \varepsilon$ gives higher priority to aggregation factor over individual preference factor. $1/v^f_i$ is the minimum of $c^p(l_{m,n})*(1+ c^{op}(l_{m,n})$ when $c^p(l_{m,n})*(1+ c^{op}(l_{m,n})) >0$. Hence $\varepsilon \ll 1/v^f_i$ makes $\varepsilon$ hardly influence the individual preference factor. Also, $\kappa \lim\limits_{x \to -\mu}f(x) \approx 1$ defines that the priority of logic factor is higher than aggregation factor. To summarize, the time logic of trips is the first thing to ensure, followed by aggregate information. On this basis, the preferred trip time slots of individuals will be followed.

\paragraph{Time period determination}

This step is to determine the time period $t^s$ based on the selected time slot $l^s$. In this dataset, for privacy reasons, we cannot fully expose an individual's real trip time.
However, it is achievable that make the trip time period distribution of generated data approximate to the real data.

Assuming $l^s=l_{a,b}$ and the current time period of $T^p$ is $t_d$, then the range of trip time period that can be selected from is $l_{r,b}$, where $r=max(a,d)$. Denote $u(t_k) = \sum_{i=1}^n(v^t_{i,k})$, and $e(t_k) = u(t_k)/\sum_{j=1}^{1440}(t_j)$. using $e^g$ to indicate the statistics of the data have been generated like $u^g$. Define $\Delta e(t_k) = e(t_k)-e^g(t_k)$, and the probability of $t_k \in l_{r,b}$ to be selected can be calculated by Eq. \ref{eq:tp_der}.

\begin{equation} \label{eq:tp_der}
    p(t^s=t_k) = \left\{
    \begin{array}{l@{\qquad}l}
        max(0,\Delta e(t_k))/\sum_{j=r}^b max(0,\Delta e(t_j)) & \sum_{j=r}^b max(0,\Delta e(t_j))>0\\
         1/|\Delta e(t_k)| / \sum_{j=r}^b (1/|\Delta e(t_j)|) & \sum_{j=r}^b max(0,\Delta e(t_j))=0\\
    \end{array}
    \qquad\quad t_k \in l_{r,b}
    \right.
\end{equation}

With the same consideration of letting $c^r$ be shared only among the same type of travellers, we propose to distinguish the traveller types when computing $\Delta e(t_k)$ to restore aggregated temporal distribution for each type of traveller.

\subsubsection*{Trip destination generation}

In the real world, trip origin and trip time are two significant elements related to the trip destination choice of individuals. Individual trip features are mainly reflected by these spatiotemporal and spatial associations of trips, which means that these information would very easily reveal the trip privacy of individuals. For privacy protection reasons, in our method, the information about the spatiotemporal association of individual trips is protected. In other words, only the trip origin (current location) is considered when determining the destination.

Note $v_{i,a,b}^{o,d}$ as the trip frequency of $v_i$ with $z_a$ and $z_b$ as the trip origin and destination, respectively. Define $v_{i,a}^{o} = \sum_m v_{i,a,m}^{o,d}$, which represents the trip frequency of $v_i$ with $z_a$ as the trip origin. Supposing the current location is $z_c$ (the origin of this trip), then the destination $z^s$ can be determined with the probability given by Eq. \ref{eq:des_der}.

\begin{equation} \label{eq:des_der}
p(z^{s}=z_t) = v^{o,d}_{i,c,t}/v^o_{i,c}     \qquad\quad z^p_i = z_c
\end{equation}

\subsubsection*{Trip path and duration generation}

Trip path refers to a sequence of spatially continuous roads (see Fig.~\ref{fig:road_zone}) by which the individual trips from the origin to the destination. There are usually multiple access paths between two traffic zones, and individuals' selections of trip paths affect road flow distribution, which is significant information for traffic condition analysis.
However, according to this study\cite{gao2019quantifying}, just a few spatiotemporal tuples can identify most individuals uniquely. Even though we are generating all trips, privacy leakage is still possible if we completely restore the individual path choices.
Thus for an individual, we sample the trip path based on its crowd (e.g., random traveller), which is a way for generalization. It can recover the flow distribution of roads and conceal individual trip preferences. Specifically, note $z_o$,$z_d$ as the origin and destination of the trip being generated, and note the set of trip paths that connect $z_o$ and $z_d$ as $P_{o,d}$. The probability to be chosen of each trip path in $P_{o,d}$ can be given by Eq. \ref{eq:path_der}.

\begin{equation} \label{eq:path_der}
p(p^{s}=p_k) = \sum_{i=1}^n v^p_{i,k}/\sum_{k}\sum_{i=1}^n v^p_{i,k}\quad\qquad p_k \in P_{o,d}
\end{equation}

Trip duration (or travel time) means the time taken to complete the trip. It is mainly related to the length of the trip path, while it is significantly affected by traffic control strategies like signal control and actual traffic condition. Since the generated trip data includes departure time, origin, destination, and trip path, the trip duration can be estimated or obtained by simulation. However, considering some users just need a possible trip duration for analysis, we give each trip's duration retrieved from the real data. Considering the correlation between trip duration and trip time slots, in determining the trip duration of a generated trip, we randomly sample among historical trips of the same trip path and trip time slot. In this way, each trip's duration was tripped by an individual in the real world with that traffic condition.

\section*{Data Records}

The city-scale synthetic individual-level vehicle trip data is released by comma-separated values (CSV) files, containing 1,829,218 trip records of 276,978 vehicle individuals in XuanCheng for one week. The fields of data record and the meanings are shown in Tab. \ref{tab:data_record1}. 
To support more applications, the road network of XuanCheng city, which matches this synthetic trip dataset, is also given and released in a Zip file. Besides, the relationship between the traffic zones and roads is released by a CSV file (see Tab. \ref{tab:data_record2} for detail). These data are available at the Figshare\cite{figshare2023} repository.

\section*{Technical Validation}

Although the trip data proposed in this paper is synthetic and a lot of effort has been made to protect individual trip privacy during generation, the dataset still has a high value for research and application. In this section, we will validate our data by comparing it with the real trip data (2019/8/12-2019/8-18) from both aggregated and individual perspectives. 

\subsection*{Aggregated level}

Aggregated level data (e.g., Fig.~\ref{fig:veh_trip_time}-\ref{fig:veh_trip_flow}) refers to the data formed by individuals' trips aggregated from spatial or temporal dimensions, such as the distribution of trips with time. 
It can be obtained from the individual-level trip data by simply counting, indicating the aggregated information of trips within the selected range. The generated data can support the aggregated level research or analysis when it keeps consistent with the real one on this level. Next, a series of comparisons of generated versus real data will be demonstrated. For quantitative evaluation, the Jensen-Shannon divergence (Eq.\ref{eq:KL}-\ref{eq:JS}) is introduced, where $p$,$q$ are the distributions based on historical data and generated data statistic, respectively. Besides, for two set $S,S'$ that satisfy $|S|=|S'|$, we denote $\frac{|S \bigcap  S'|}{|S|}$ as the overlap ratio of these two sets, which will be used in spatial dimension evaluation.

\begin{equation} \label{eq:KL}
    D_{KL}(p||q) = \sum_{i=1}^N p(x_i)\cdot log\frac{p(x_i)}{q(x_i)}
\end{equation}

\begin{equation} \label{eq:JS}
    D_{JS}(p||q) = \frac{1}{2}D_{KL}(p||\frac{p+q}{2})+\frac{1}{2}D_{KL}(q||\frac{p+q}{2})
\end{equation}

\paragraph{Temporal dimension}

\begin{figure}[H]
\centering
\subfigure{\includegraphics[width=0.98\textwidth]{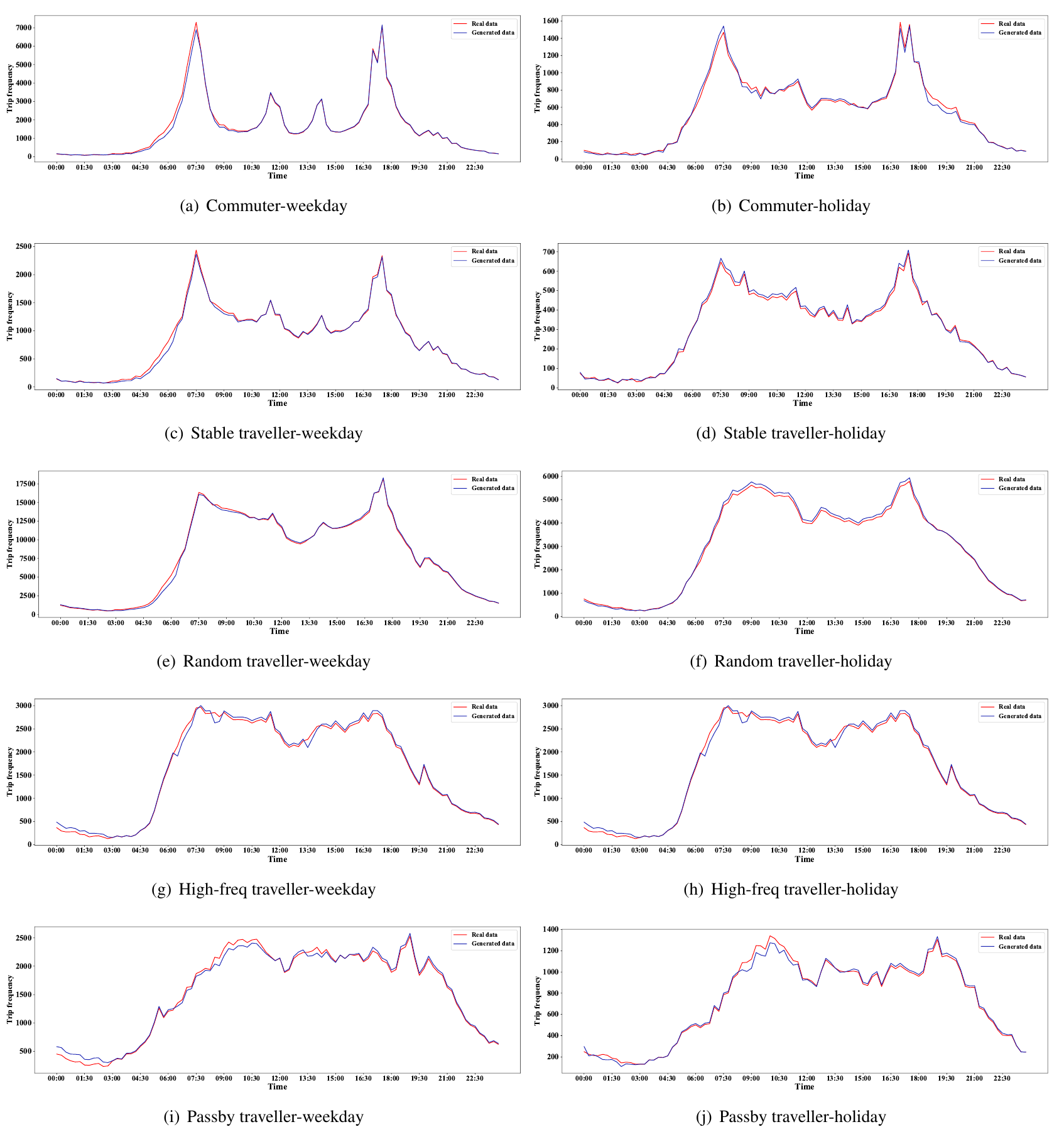}}
\caption{Distributions of trip time of the synthetic data and the real data (Statistics and plots at 15-minute granularity).}
\label{fig:agg_time}
\end{figure}

The distributions of trips with time (distinguished weekday and holiday) of the synthetic and real data are shown in Fig. \ref{fig:agg_time}, and the Jensen-Shannon divergences of the two distributions are shown in Table \ref{tab:agg_Jstime}. The vertical axis of Fig. \ref{fig:agg_time} adopts the frequency of trips to show that the quantity of generated trips is also similar to the real one. Moreover, the result of high consistency would be kept when considering a smaller time scale, like each day or a specific time slot.

\paragraph{Spatial dimension}

The spatial information of urban trips can be portrayed from three levels: (1) the visited hotness of traffic zones; (2) the dominated origin-destination of trips; (3) the distribution of road access frequency.

\begin{figure}[ht]
\centering
\subfigure{\includegraphics[width=0.98\textwidth]{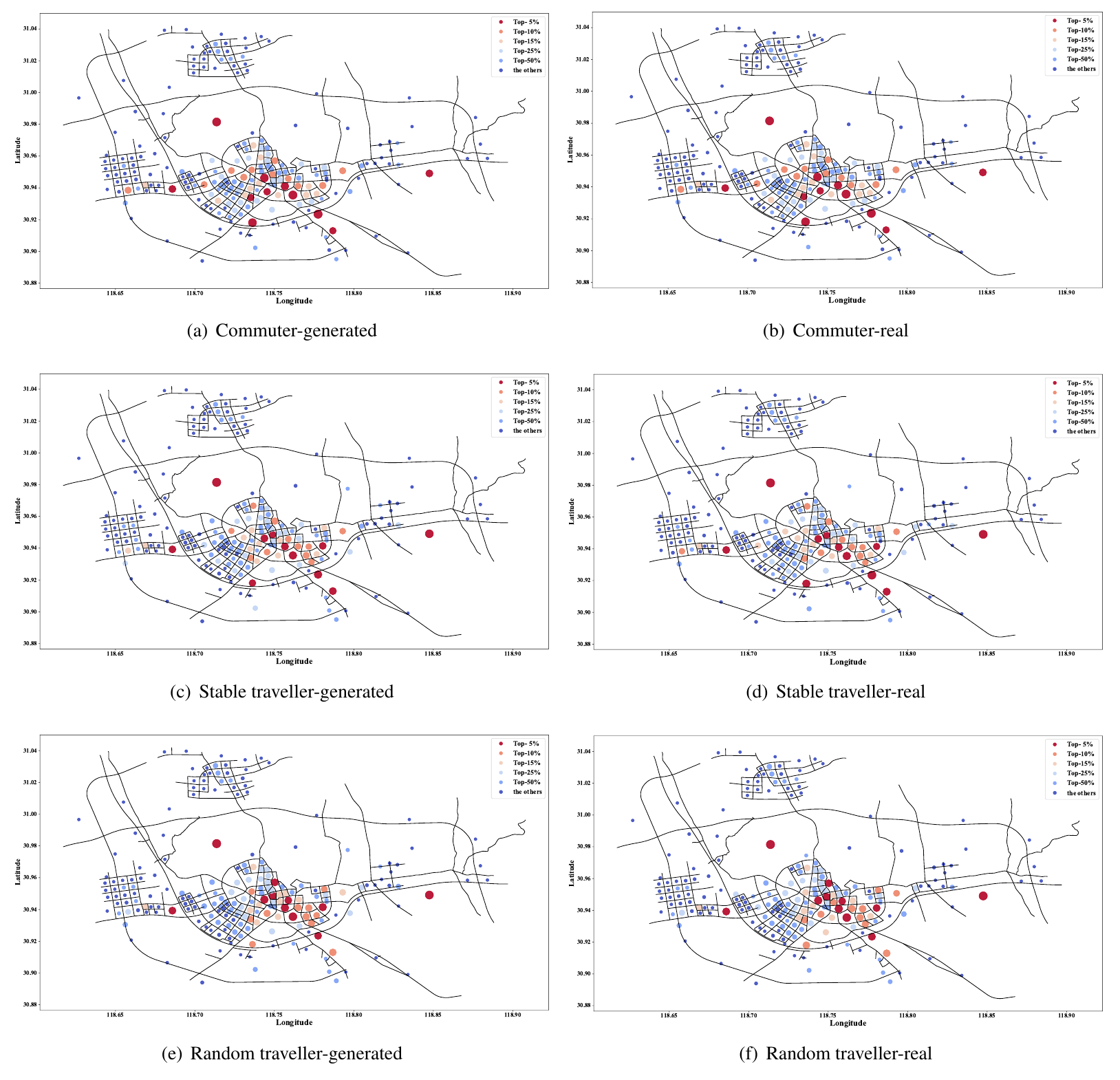}}

\caption{Distribution of visited hotness of traffic zones on the road network (Commuter and other two types of traveller).}
\label{fig:spa_hotness1}
\end{figure}

\begin{figure}[ht]
\centering
\subfigure{\includegraphics[width=0.98\textwidth]{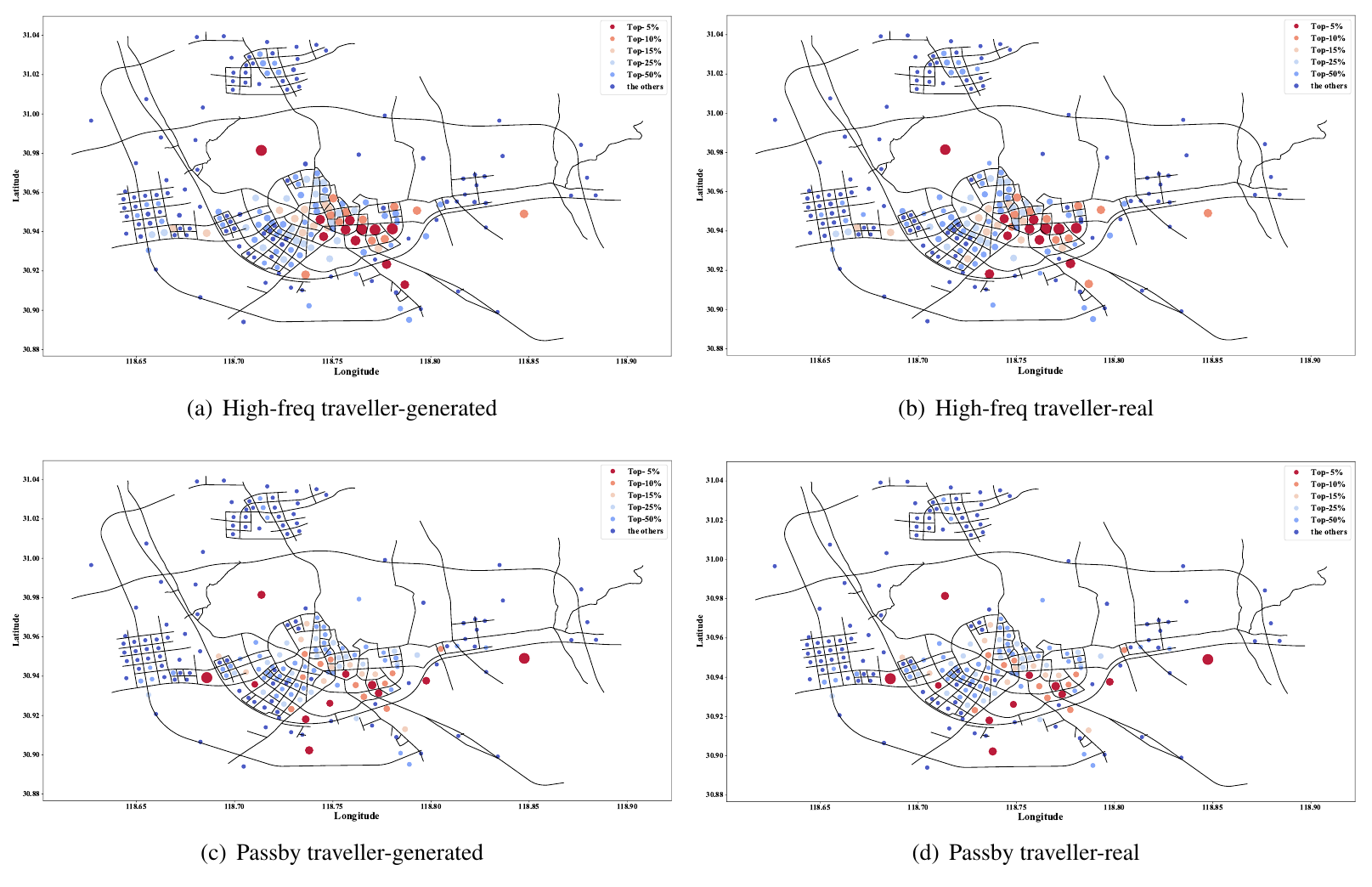}}

\caption{Distribution of visited hotness of traffic zones on the road network (High-freq and passby travellers).}
\label{fig:spa_hotness2}
\end{figure}

The visited hotness of traffic zones can reveal the main activity areas of travellers in the city.
We have counted the top-$k$ most frequently visited traffic zones of the synthetic and real dataset with different $k$ values. The overlap ratio is calculated and formed Tab. \ref{tab:agg_hot}. 
The hotness of traffic zones may vary over time, so we conducted further experiments, such as limiting the time to specific periods. The results show that the performance shown in Tab. \ref{tab:agg_hot} is stable. Further, Fig.~\ref{fig:spa_hotness1}-\ref{fig:spa_hotness2} show the distributions of traffic zones with different hotness levels on the road network. In addition to using colour to distinguish the hotness levels, a larger dot indicates a higher visited frequency, i.e., the largest red dot is the most visited traffic zone.

From the perspective of origin-destination of trips, we count the trip frequency of different origin-destination of the synthetic and real data. The overlap ratios of top-$k$ dominated origin-destinations are shown in Tab. \ref{tab:agg_od}. The results can be maintained when distinguishing weekdays and holidays.

\begin{figure}[ht]
\centering
\subfigure{\includegraphics[width=0.98\textwidth]{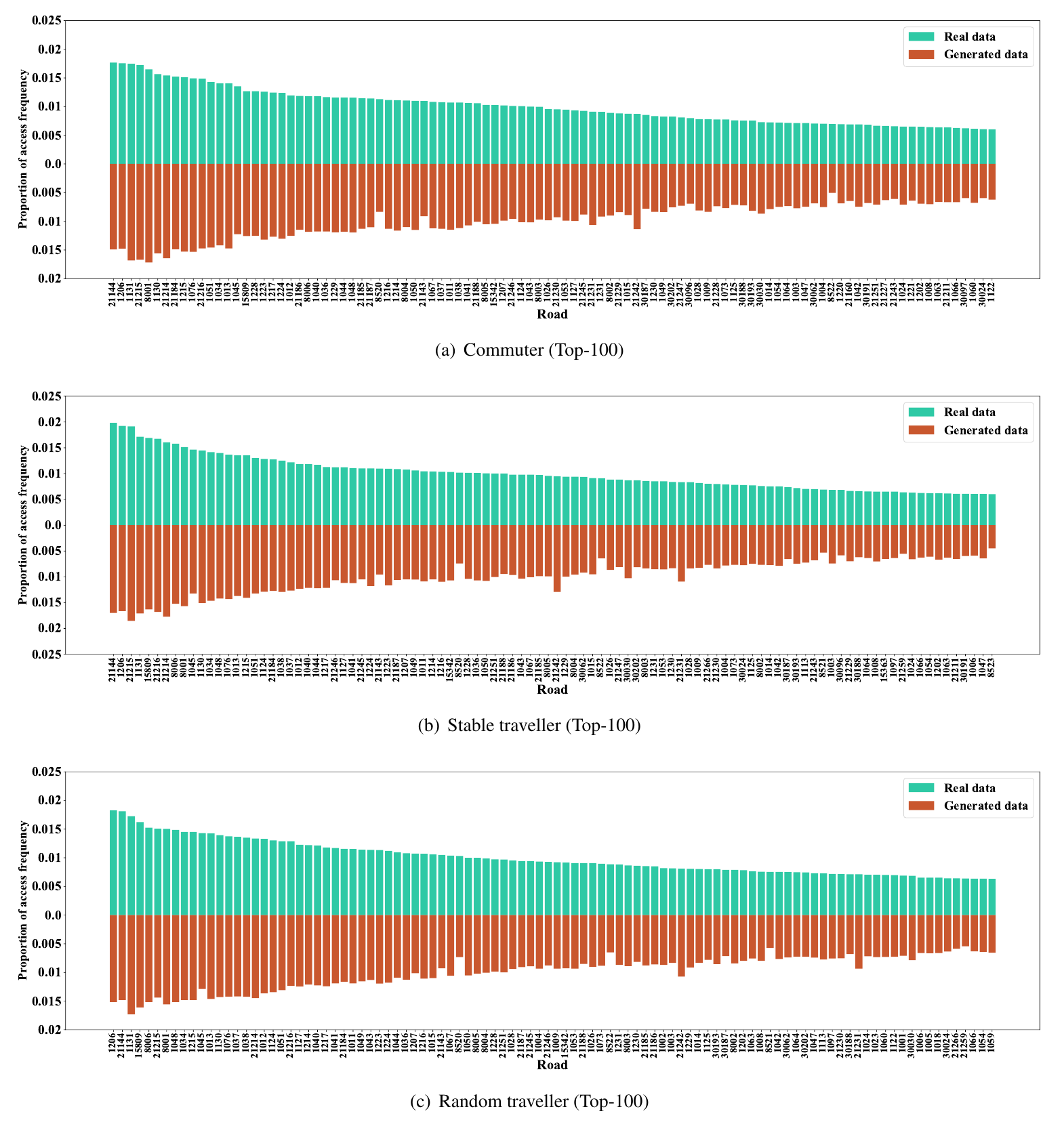}}
\caption{The access frequency of major roads of the synthetic and real data (Commuter and other two types of traveller).}
\label{fig:agg_flow}
\end{figure}

\begin{figure}[ht]
\centering
\subfigure{\includegraphics[width=0.98\textwidth]{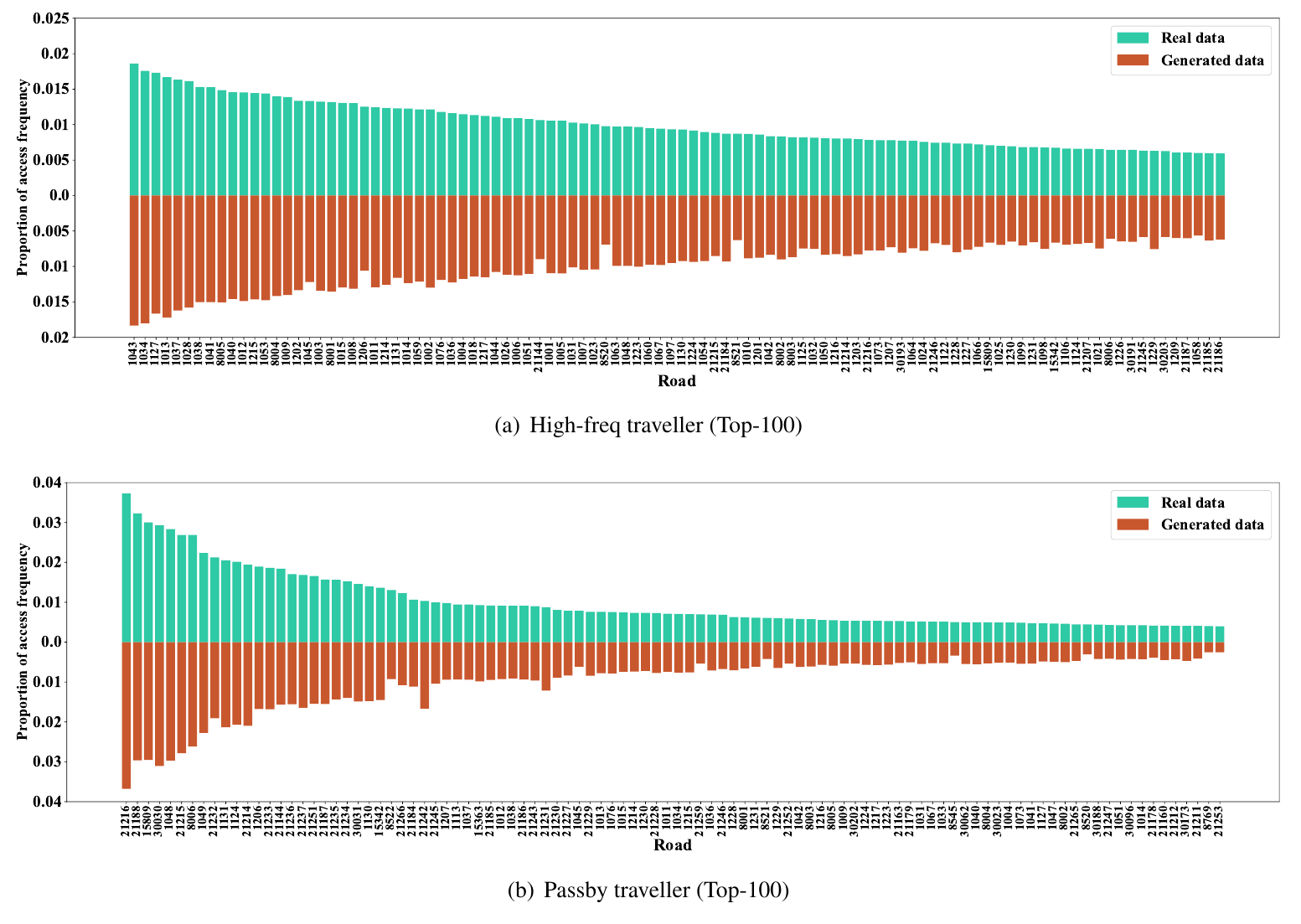}}
\caption{The access frequency of major roads of the synthetic and real data (High-freq and passby travellers).}
\label{fig:agg_flow2}
\end{figure}

\begin{figure}[ht]
\centering
\subfigure{\includegraphics[width=0.98\textwidth]{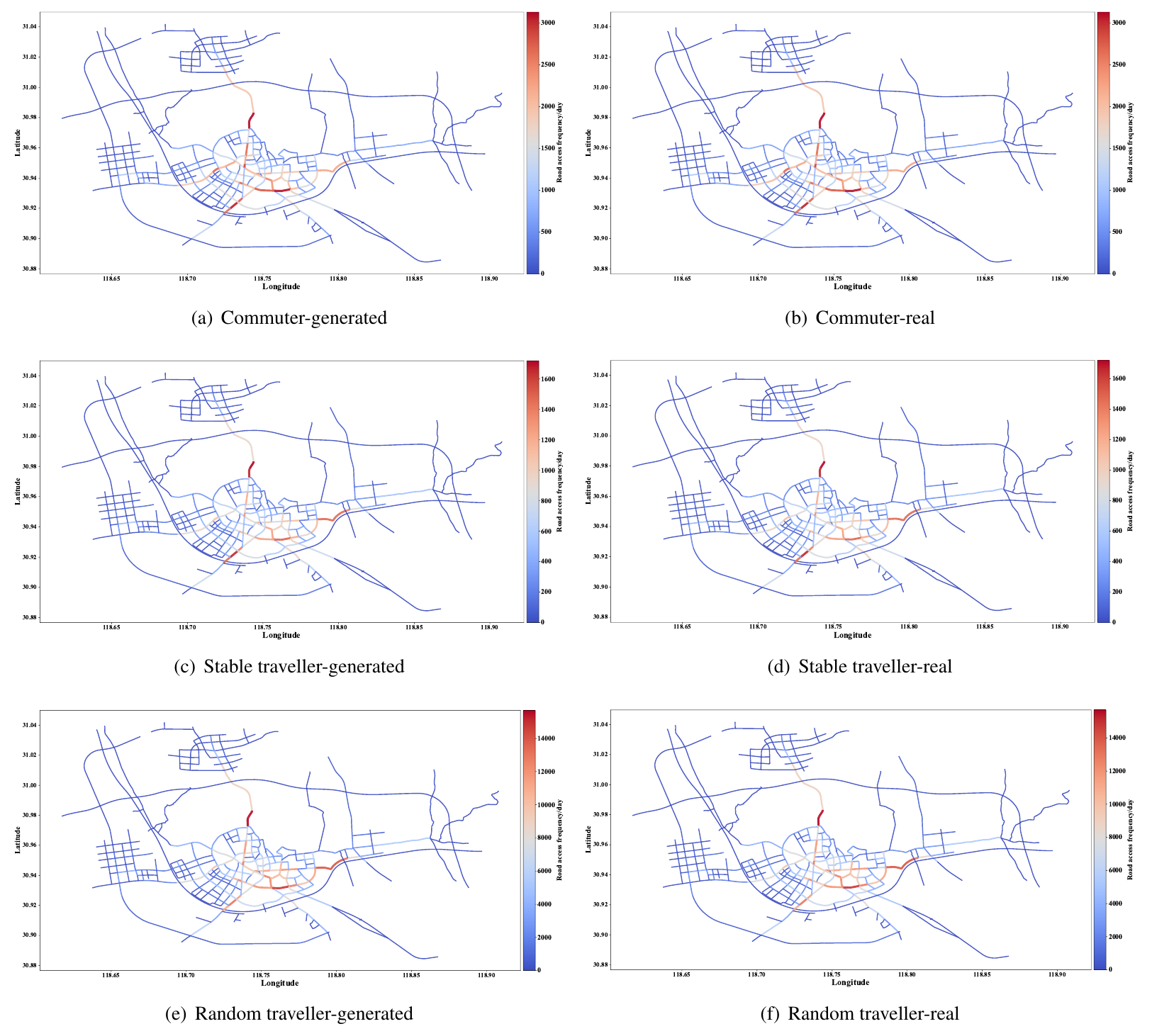}}

\caption{Distribution of road access frequency on the road network (Commuter and other two types of traveller).}
\label{fig:spa_flow1}
\end{figure}

\begin{figure}[ht]
\centering
\subfigure{\includegraphics[width=0.98\textwidth]{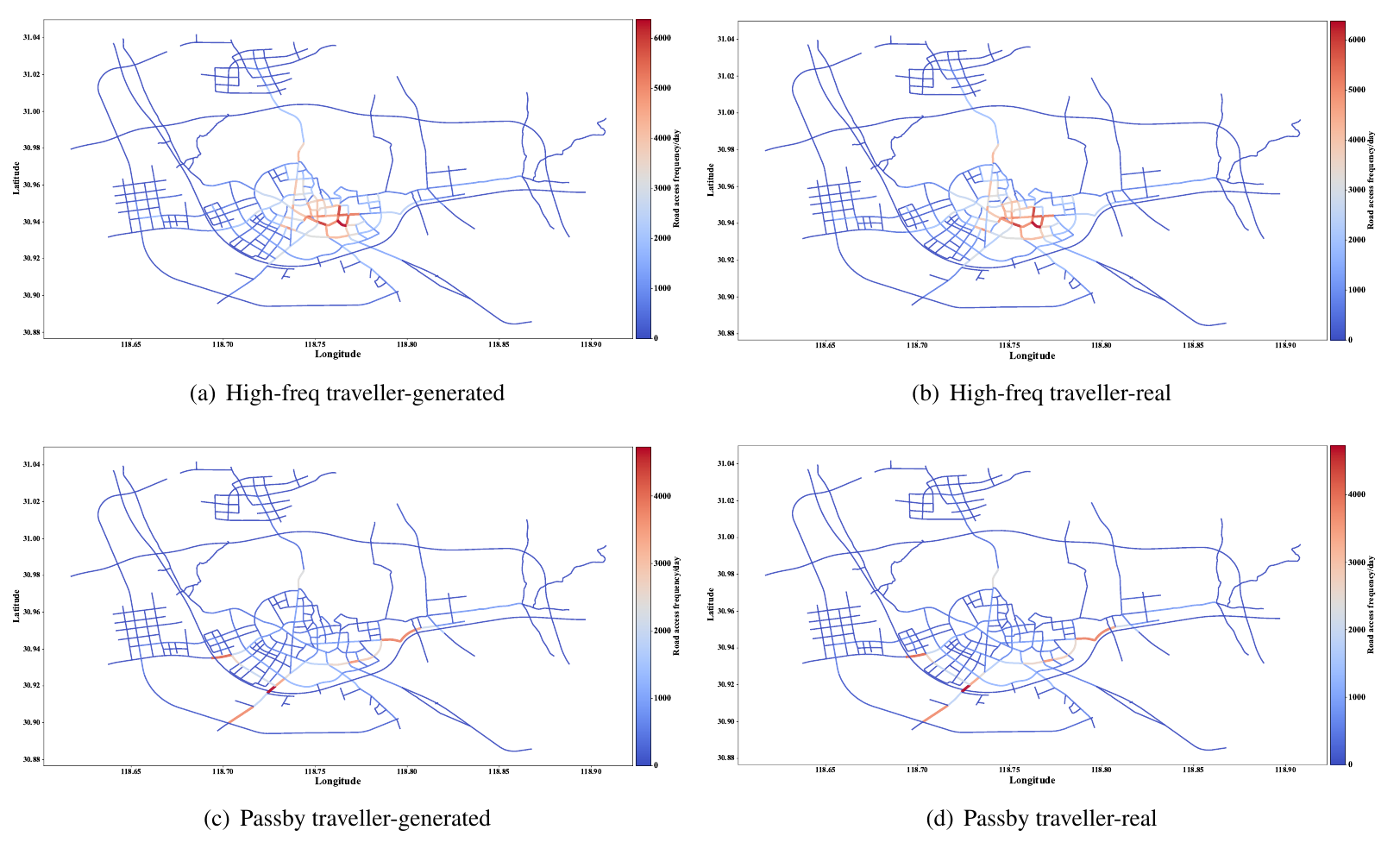}}

\caption{Distribution of road access frequency on the road network (High-freq and passby travellers).}
\label{fig:spa_flow2}
\end{figure}

The distribution of access frequency of roads can help identify critical roads and is valuable for urban transportation management. 
The access frequency proportion of major roads of the synthetic and real data are shown in Fig. \ref{fig:agg_flow}-\ref{fig:agg_flow2}. In addition, the Jensen-Shannon divergence of the two distributions is calculated and shown in Fig.~\ref{tab:road_js}. When limiting the time range to specific slots, such as the morning and evening rush hours, the Jensen-Shannon divergence will be double but still at a pretty low level. Further, the distributions of daily access frequency of roads on the road network are shown in Fig.~\ref{fig:spa_flow1}-\ref{fig:spa_flow2}.

In summary, the generated data can restore the aggregated level information of the real data. Besides, we found the bias between the generated and real data is close to that of two weeks of the real data. This means that the generated data will not be distinguished by aggregated information.

\subsection*{Individual level}

The availability of the generated trip data on the individual level is based on the reasonableness of trips from the individual perspective. 
Specifically, it includes two levels of information:  1) the reasonableness of a single individual's trips; 2) distributions of individual-based statistics.
Next, we will validate the generated data from these two levels. 

\paragraph{Reasonableness of single individual trips}

In the real world, individual travel follows certain laws. For instance, Individuals' trips are spatially continuous, and there is generally an interval of time between trips. Therefore, the trip data of an individual that follows these laws are considered reasonable. Next, we will explain the reasonableness of the generated trip data in the following aspects. 

\begin{itemize}
    \item \textbf{Trip frequency}. 
    The trip frequency of an individual is generally in a reasonable range.
    In the generation, the daily trip frequency is determined with an individual in the real world as the template. In this case, the generated individual's trip frequency each day, as well as the cumulative frequency of trips in a week, is in a reasonable range. 
    \item \textbf{Trip time interval.} 
    There is a certain time interval between two consecutive trips of an individual in the real world.
    Thanks to the introduction of trip time logic and preference factors in trip time determination, there are almost no two trips with very short intervals (e.g., a few seconds) in the generated data, which are considered abnormal and should be merged.
    \item \textbf{Trip spatial range.} When determining the destinations for generating trips, we take the destinations visited by a real individual as the candidate set. Thus the trip spatial range of the generated individual will not exceed that of its template individual in reality, which makes the trip spatial range of all individuals in the synthetic dataset reasonable.
    \item \textbf{Spatial continuity of trips.} 
    Objectively, the trips of the individuals should all be of spatial continuity (the destination of the previous trip is the same as the origin of the next trip). However, it cannot be completely ensured due to incorrect license plate recognition and driving out of the perception boundary. In the synthetic data, the trip spatial continuity ratios of commuters, stable travellers, and random travellers are 81.68\%, 79.55\%, and 76.96\%. Compared to the real data, they improved by 10.86\%, 12.55\%, and 11.82\%, respectively.
\end{itemize}

\paragraph{Distributions of individual-based statistics}

Although the trips are reasonable from an individual perspective, the availability of the generated data on the individual level would be decreased if the distribution of individual trip characteristics does not match a real city. 
In this section, we focus on the following three distributions of individual-based statistics.

\begin{figure}[ht]
\centering
\subfigure{\includegraphics[width=0.98\textwidth]{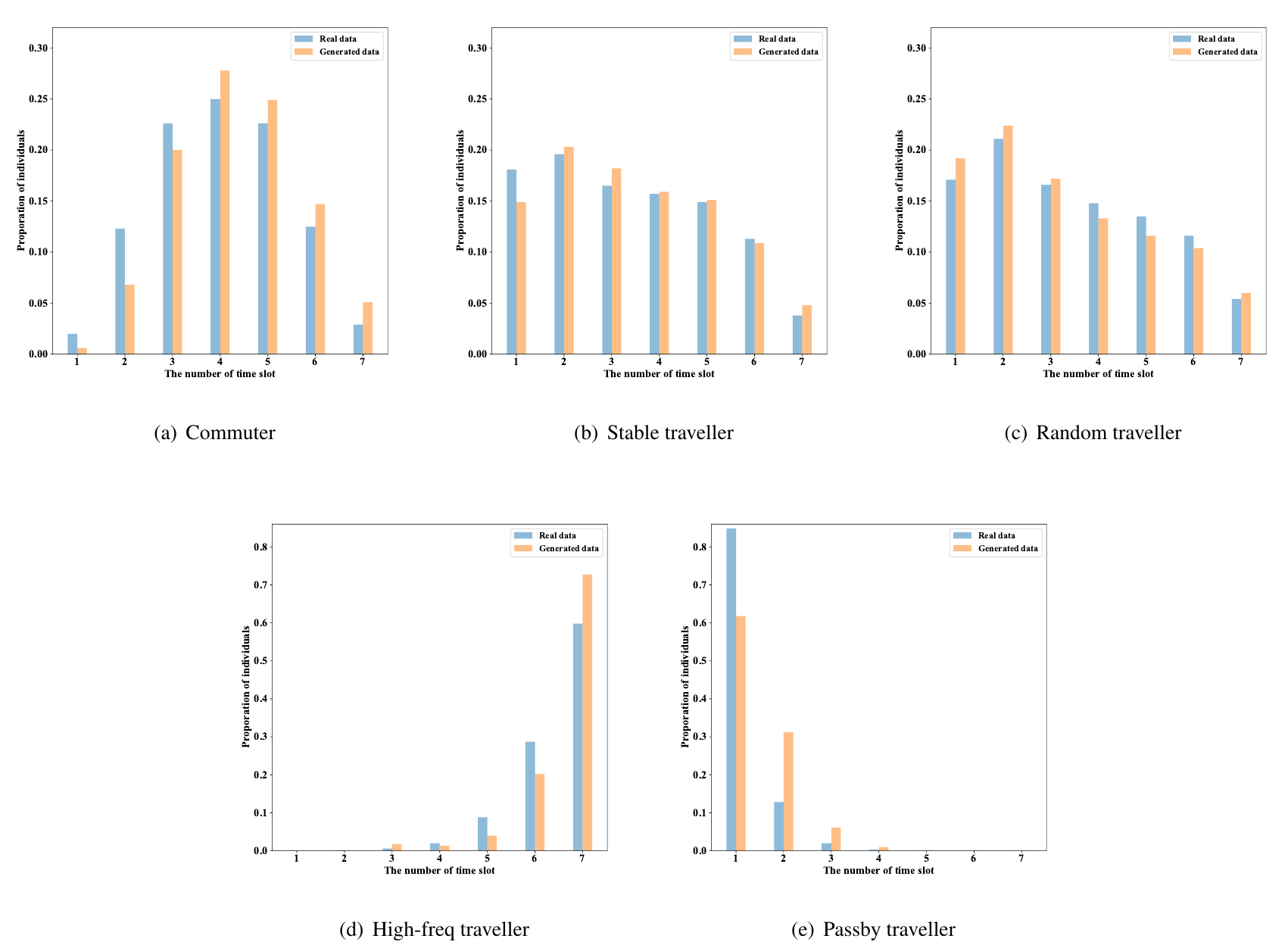}}
\caption{Distribution of the number of individual trip time slots.}
\label{fig:veh_trip_sd}
\end{figure}

\begin{figure}[ht]
\centering
\subfigure{\includegraphics[width=0.88\textwidth]{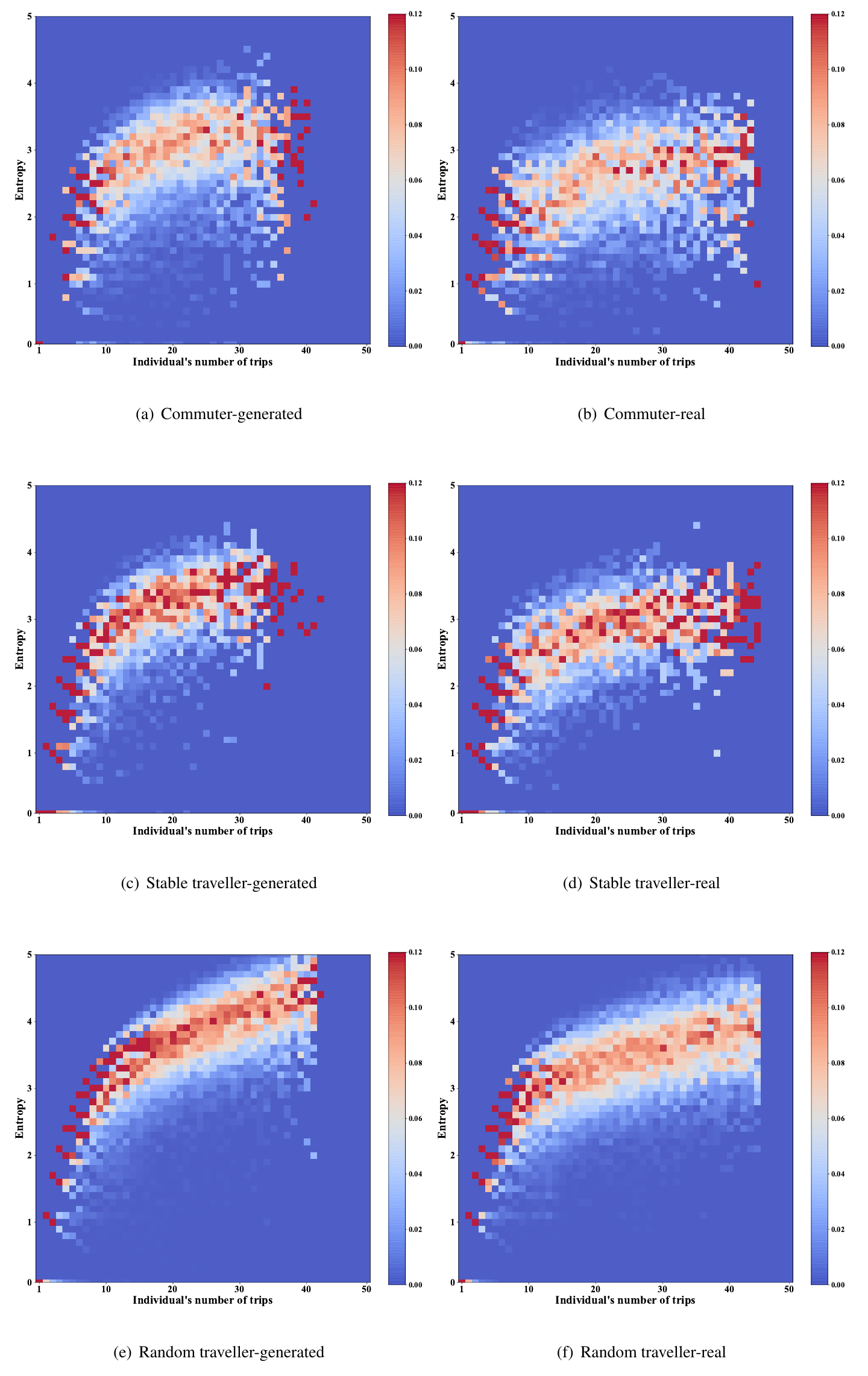}}

\caption{Distribution of entropy of individual trip destinations (Commuter and other two types of traveller).}
\label{fig:entropy1}
\end{figure}

\begin{figure}[ht]
\centering
\subfigure{\includegraphics[width=0.88\textwidth]{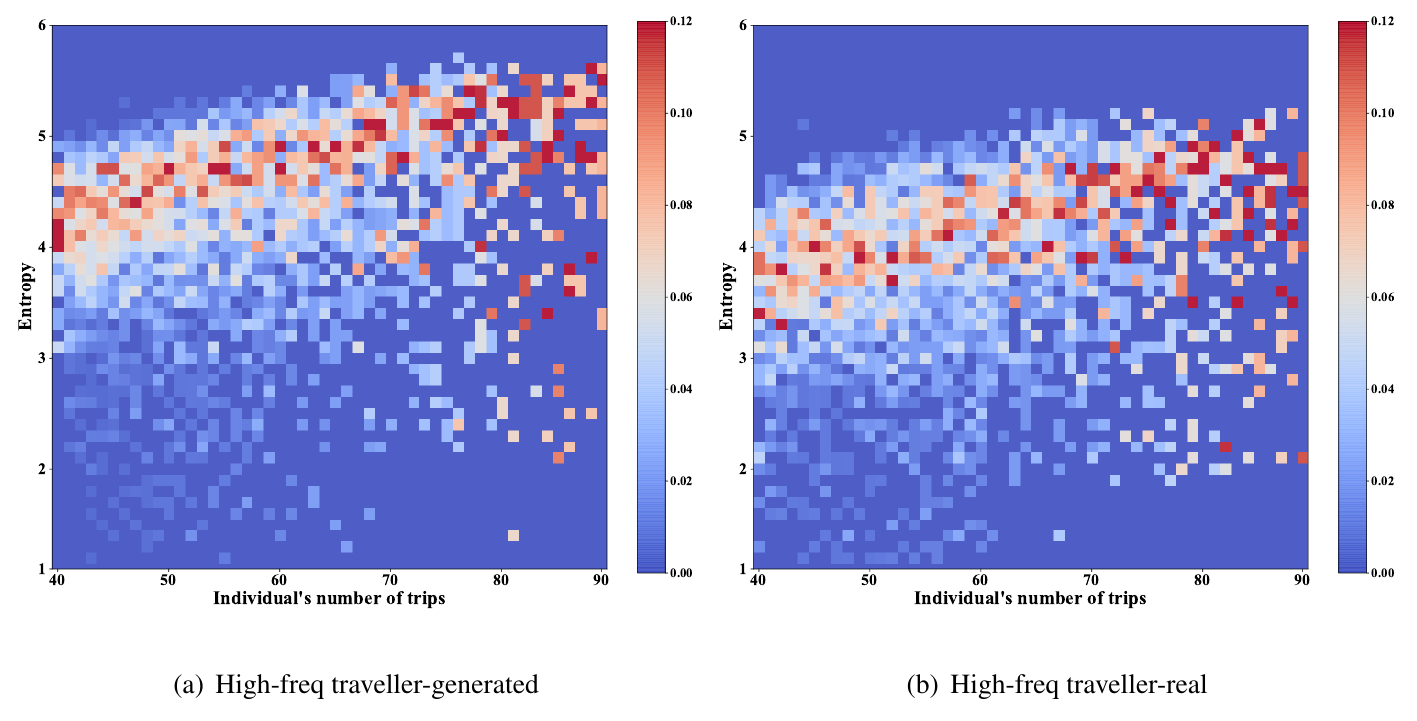}}
\caption{Distribution of entropy of individual trip destinations (High-freq traveller).}
\label{fig:entropy2}
\end{figure}

\begin{itemize}
    \item \textbf{Distribution of trip frequency.} The daily trip frequency of individuals is calculated based on the historical trips of real individuals. Thus, the distribution of individual trip frequency of the generated data is highly consistent with the real data.
    \item \textbf{Distribution of the number of trip time slots.} The number of time slots covered by individuals in the generated data and the real data of one week are counted and displayed in Fig.~\ref{fig:veh_trip_sd}.
    \item \textbf{Distribution of entropy of trip destinations.} Entropy is a measure of the regularity of travellers \cite{cheng2021probabilistic}. The distributions of the generated and real data are shown in Fig.~\ref{fig:entropy1}-\ref{fig:entropy2} (the trip frequency of passerby travellers is too low to analyze entropy). The colour blocks represent the proportion of individuals among all individuals with the same number of trips. It should be noted that the dynamic range of colours is set to $0-0.12$, and the scale higher than $0.12$ is also marked with the same colour as $0.12$ (i.e., red), for better display of details. Since the destination candidate set for generation considers the individual's destination choice for one month while comparing the real data with one week, the entropy of individuals in the generated data is slightly higher than that of the real data. This reflects our data privacy protection and has little impact on data availability.
\end{itemize}

\section*{Usage Notes} \label{sec:usage}

All datasets open in this paper are in file form, and users can access them in their entirety without any further permission. To make the individual-level trip dataset publicly available, we perform works on traveller's trip privacy protection that we have mentioned in the Methods section, which inevitably affects the usability of the data. Here, to prevent the data's misuse, we would like to remind potential users of the characteristics of the synthetic (or generated) trip dataset and the tasks for which it is unsuitable. First, The average daily trip frequency for each individual in the generated trip dataset is reasonable, and the overall distribution is realistic. However, for a single individual, the distribution of individual trip frequencies in the synthetic dataset is more uniform. Therefore, we do not recommend using this synthetic dataset to analyse the variation and patterns of individual single-day trip frequency. Second, in the synthetic dataset, traveller trip temporal and spatial preferences are reliable, but trip spatio-temporal associations are broken. Hence, this dataset is not applicable for tasks involving the analysis of trip spatio-temporal associations of travellers. 

In addition to reminding synthetic data of the limitations of its use, we would also like to make a few notes to facilitate better data usage. First, the traveller labels given in the dataset are not true values and can be reassigned or divided by the user according to the specific task. Second, the individuals whose ``traveller\_ID" starts with ``Wan\_P" can be considered as local vehicles (or travellers). Thus this synthetic dataset can support the trip pattern analysis for local and foreign vehicles. Third, we may have prior knowledge or common sense about travelling in the city, such as commuters usually travel to their workplace in the morning. Most travellers may behave in line with common sense, but we also need to be aware of the presence of unusual travellers. For example, in our previous analysis, we found that there were night commuters in the city who travelled to their homes in the morning.

\section*{Code availability}

The codes for trip generation algorithms and the synthetic dataset validation are available via the GitHub repositories\cite{github2022}.


\section*{Acknowledgements}

This research was supported by the National Natural Science Foundation of China (No. U1811463 and No. U21B2090).

\section*{Author contributions statement}
G.L. developed the theoretical framework and performed the computations. 
Y.C contributed to the technical details of the the theory.
Y.W contributed to part of the experiments.
P.N contributed to text proofreading.
Z.Y. contributed to the original data. 
Z.H. supervised the findings of this work.
All authors discussed the results and contributed to the final manuscript.


\section*{Competing interests}

The authors declare no competing interests.

\section*{Figures \& Tables}

\begin{table}[H]
\centering
\begin{tabular}{|l|l|}
\hline
Notation & Description \\
\hline 
$v^f_i$ & The trip frequency of $v_i$ \\ 
\hline 
$v^t_{i,k}$ & The trip frequency of $v_i$ with $t_k$ as departure time.\\ 
\hline 
$v^o_{i,a}$ & The trip frequency of $v_i$ with $z_a$ as the origin.\\ 
\hline
$v^d_{i,b}$ & The trip frequency of $v_i$ with $z_b$ as the destination. \\ 
\hline
$v^p_{i,k}$ & The number of trips of $v_i$ with trip path $p_k$. \\
\hline
\end{tabular}
\caption{\label{tab:notations} Description of some numeric variables.}
\end{table}

\begin{table}[h]
\flushleft
\begin{tabular}{|l|l|}
\hline
Column name & Description \\
\hline
traveller\_ID & The identify of the individual\\ 
\hline 
traveller\_type & The traveller type of the individual \\ 
\hline 
Date & The date when the trip happened\\ 
\hline 
Departure\_time & The departure time (minute-level) of the trip \\ 
\hline 
Time\_slot & The time slot that the departure time belongs to \\ 
\hline 
O\_zone & The origin of the trip, represented by traffic zone\\ 
\hline 
D\_zone & The destination of the trip, represented by traffic zone\\ 
\hline
Path & The path that the trip take (roads are separated by ``-")  \\ 
\hline 
Duration & The length of time for completing the trip \\
\hline
\end{tabular}
\caption{\label{tab:data_record1} The synthetic (or generated) individual-level trip data attributes.}
\end{table}

\begin{table}[h]
\flushleft
\begin{tabular}{|l|l|}
\hline
Column name & Description \\
\hline
Zone\_ID & The ID of the traffic zone\\ 
\hline 
Longitude & The center point longitude of the traffic zone \\ 
\hline 
Latitude & The center point latitude of the traffic zone \\ 
\hline 
Roads & The related roads of the traffic zone\\ 
\hline 
\end{tabular}
\caption{\label{tab:data_record2} Attributes of data about the relationship of traffic zones and roads.}
\end{table}

\begin{table}[H]
\centering
\caption{Jensen-Shannon divergences of trip frequency with time (generated data vs. real data).}
\begin{tabular}{|c|c|c|c|c|c|}
\hline
        & Commuter & Stable traveller & Random traveller & High-freq   traveller & Passby traveller \\ \hline
Weekday & 0.000500 & 0.000304        & 0.0002099       & 0.000307            & 0.0004044       \\ \hline
Holiday & 0.000364 & 0.000100        & 0.000100       & 0.000196            & 0.0002049       \\ \hline
\end{tabular}
\label{tab:agg_Jstime}
\end{table}

\begin{table}[ht]
\centering
\caption{Degree of overlap of hot traffic zones.}
\begin{tabular}{|c|c|c|c|c|c|}
\hline
        & Commuter & Stable traveller & Random traveller & High-freq traveller & Passby traveller \\ \hline
Top-5\%  & 90\%      & 100\%            & 90\%              & 90\%                & 100\%                \\ \hline
Top-10\%  & 95\%      & 100\%            & 90\%              & 90\%                & 95\%             \\ \hline
Top-15\%  & 93\%      & 97\%             & 97\%             & 93\%               & 100\%                \\ \hline
Top-25\%  & 96\%      & 96\%             & 98\%             & 98\%               & 100\%                \\ \hline
Top-50\% & 98\%      & 96\%             & 97\%             & 97\%               & 100\%                \\ \hline
\end{tabular}
\label{tab:agg_hot}
\end{table}

\begin{table}[ht]
\centering
\caption{Degree of overlap of main OD combinations.}
\begin{tabular}{|c|c|c|c|c|c|}
\hline
         & Commuter & Stable traveller & Random traveller & High-freq traveller & Passby traveller \\ \hline
Top-1\%  & 74.0\%   & 79.4\%          & 81.2\%          & 80.4\%            & 93.3\%          \\ \hline
Top-5\%  & 80.6\%   & 80.4\%          & 84.9\%          & 78.0\%            & 93.5\%          \\ \hline
Top-10\% & 79.6\%   & 81.0\%          & 86.8\%          & 77.5\%            & 92.0\%          \\ \hline
Top-30\% & 82.6\%   & 79.7\%          & 88.5\%          & 79.1\%            & 85.2\%          \\ \hline
Top-50\% & 82.2\%   & 78.4\%          & 87.2\%          & 81.0\%            & 80.3\%          \\ \hline
\end{tabular}
\label{tab:agg_od}
\end{table}

\begin{table}[ht]
\centering
\caption{Jensen-Shannon divergences of the access frequency of main roads (generated data vs. real data).}
\begin{tabular}{|c|c|c|c|c|}
\hline
 Commuter & Stable traveller & Random traveller & High-freq   traveller & Passby traveller \\ \hline
0.00074 & 0.00095        & 0.00083       & 0.00052            & 0.0016       \\ \hline
\end{tabular}
\label{tab:road_js}
\end{table}

\end{document}